\documentclass[12pt,preprint]{aastex}

\def\kms{\ifmmode {\rm \ km \ s^{-1}}\else \rm km \ s^{-1}\fi}
\def\cm{\ifmmode {\rm \ cm }\else $\rm cm$\fi}

\received{17 October 2000}
\accepted{.. .. 2000}

\begin{document}

\title{Ly$\alpha$ line formation in starbursting galaxies \\
II. Extremely Thick, Dustless, and Static HI Media}

\author{Sang-Hyeon Ahn}
\affil{Korea Institute for Advanced Study, 207-43 Cheongyangri-dong, Dongdaemun-gu, Seoul, 130-012, Korea}
\email{sha@kias.re.kr}

\author{Hee-Won Lee}
\affil{Department of Earth Science, Sejong University, Seoul 143-747, Korea}

\and

\author{Hyung Mok Lee}
\affil{School of Earth and Environmental Sciences, Astronomy Program, Seoul National University, Seoul 151-742, Korea.}

\begin{abstract}
The Ly$\alpha$ line transfer in an extremely thick medium of neutral hydrogen
is investigated by adopting an accelerating scheme in our Monte Carlo code
to skip a large number of core or resonant scatterings. This scheme reduces
computing time significantly with no sacrifice in the accuracy of the results.
We applied this numerical method to the Ly$\alpha$ transfer 
in a static, uniform, dustless, and plane-parallel medium.
Two types of photon sources have been 
considered, the midplane source and the uniformly distributed sources.
The emergent profiles show double peaks and absorption
trough at the line-center. We compared our results with the analytic 
solutions derived by previous researchers, and confirmed that
both solutions are in good agreement with each other.
We investigated the directionality of the emergent Ly$\alpha$ photons 
and found that limb brightening is observed in slightly thick media
while limb darkening appears in extremely thick media.  
The behavior of the directionality is noted to  
follow that of the Thomson scattered radiation in electron clouds,
because both Ly$\alpha$ wing scattering and 
Thomson scattering share the same Rayleigh scattering phase function.
The mean number of wing scatterings just before escape is in exact
agreement with the prediction of the diffusion approximation.
The Ly$\alpha$ photons constituting the inner part of the
emergent profiles follow the relationship derived from the diffusion
approximation. We present a brief discussion on the application
of our results to the 
formation of Ly$\alpha$ broad absorption troughs and P-Cygni type 
Ly$\alpha$ profiles seen in the UV spectra of starburst galaxies. 
\end{abstract}
\keywords{line: formation --- radiative transfer --- galaxies: starburst --- galaxy: formation}

\section{Introduction}
Recent observational achievements in spectroscopy of
galaxies in the very early universe \citep{ste96,ste99}
have enabled us to study starburst galaxies whose star 
formation rates are very high. Since these galaxies are at $3<z<5$,
we can observe their ultraviolet continua and Ly$\alpha$
in the optical band, which can be covered by ground-based 
large telescopes. Furthermore, Ly$\alpha$ is one of the most
conspicuous emission lines, and the investigation of 
the Ly$\alpha$ line transfer becomes important 
in understanding the physical nature of these galaxies.

The Ly$\alpha$ profiles of many starburst galaxies in the early universe
are categorized into three types : symmetric Ly$\alpha$ emission,
asymmetric or P-Cygni type one, and broad absorption 
in damping wings \citep{ten99}.
Moreover, local starburst galaxies also show similar
characteristics \citep{kun98}. The P-Cygni type Ly$\alpha$ is believed to
originate from the expanding neutral medium surrounding
the Ly$\alpha$ source, which is supported by the existence of the
low ionization interstellar absorption lines blueshifted relative
to the line-center \citep{kun98,lee98,all00}.
The Voigt profile fittings for the broad absorption lines extending
to the Lorentzian wings and P-Cygni type absorption lines revealed that
their neutral hydrogen column density $N_{\rm HI}=10^{19}-10^{22}\ {\rm cm^{-2}}$
or the line-center optical depth $\tau_0=10^6-10^9$.
However, the Voigt profile fitting is not satisfactory
in the red part of Ly$\alpha$ emission which is thought to be
contributed by back-scattered photons.
This is exemplified by the ill-fitted red part of
Ly$\alpha$ emission in the spectrum of Haro 2 \citep{leg97}.
No adequate computational method to describe the back-scattered photons
has been developed so far.

In extremely thick media of hydrogen, where Ly$\alpha$ photons wander
in both real space and frequency space, the line transfer can be described by
the diffusion approximation. In the diffusion approximation 
we assume that only wing scatterings play a significant role 
in transferring Ly$\alpha$ photons in space.
However, this approximation does not fully take into account the 
fact that the photons move back and forth between core and wing
wavelength regimes during their transfer in an extremely thick medium.
While being a core photon
the scatterings are resonant and therefore local. Wing scatterings
are associated with significant spatial excursion. Furthermore, the scattering
phase functions for resonant scatterings associated with
the $S_{1\over2}\rightarrow P_{{1\over2}, {3\over2}}$ transitions 
are different from that
associated with wing scatterings which is idential with that of the classical
Rayleigh scattering.
This fact should be considered in a very careful manner in computing
the physical quantities such as the polarization and the
angular distribution of the emergent radiation field.

In this paper, we investigate the Ly$\alpha$ line transfer
in an extremely thick and static medium by using a Monte Carlo method.
We will focus only on the dustless case in this work,
even though the dusty case is more realistic. The effect
of dust was briefly dealt by \citet{all00}
and more careful investigation is deferred.
Our assumption can be valid for the primeval objects
that had not experienced any star-formation episode.

This is the second paper in a series on the Ly$\alpha$ profile 
formation in starbursting galaxies. In the first 
paper \citep{all01} we studied the Ly$\alpha$ transfer 
in moderately thick media. We use the Monte Carlo method, which
is frequently adopted in the Ly$\alpha$ line transfer,
because of its simplicity and flexibility.
However, the Monte Carlo method becomes inefficient when the 
line-center optical depth $\tau_0$ of scattering medium is very large.
We need to follow the scattering of each photon until the escape, and
the number of scattering is proportional to $\tau_0$ for very
thick case. This means that the number of computations grows in
proportional to $\tau_0$ which could vary by several orders of
magnitudes depending on the situation. Our previous version of the
code consumes unacceptably long computing time for extremely thick cases.
We therefore extend our efforts 
to very large optical depth regime in this paper by
developing the `accelerating scheme'.

This paper is composed as follows.
In section 2, we describe our scheme to accelerate the Monte Carlo code.
In section 3 we show our results. In section 4 we discuss possible
applications of our code. The final section summarizes our major
findings.

\section{Configuration and Monte Carlo Method}
Since we are interested in the emergent Ly$\alpha$ profiles in various
situations, we first introduce the dimensionless parameter $x$ defined by
\begin{equation}
x \equiv \Delta\nu/\Delta\nu_D=(\nu-\nu_0)/\Delta\nu_D,
\end{equation}
which describes the frequency shift from the line-center $\nu_0$ in units
of the Doppler shift $\Delta\nu_D\equiv \nu_0 (v_{th}/c)$.
Here $v_{th}$ is the thermal speed of the scattering medium,
$c$ is the speed of light, and $\nu_0$ is the line-center frequency.\\
\indent
The medium considered in this study is a plane-parallel slab,
which is uniformly filled with pure hydrogen atoms.
The optical depth at the Ly$\alpha$ line-center 
from the center to both sides of the medium
along the normal direction is $\tau_0$.
We consider two types of the source distribution: a source in the
midplane, and sources uniformly distributed along the vertical line.
We show the configuration considered in this study in Figure~1.
The underlying ideas for these two cases is that the midplane source
can model a giant H~II region embedded in a neutral region while
the uniformly distributed source can mimic the starbursting region
which has young stars mixed with the partially ionized gas.
These configurations are the same ones that have been considered
in the previous works \citep{neu90,ada72}.\\
\indent
There have been several Monte Carlo approaches to the resonance line 
transfer in an optically thick and static medium in the literature
\citep{ave68,ada72,mei81,gou96}.
The main advantages of Monte Carlo methods lie in
the geometrical flexibility of scattering media 
and the simplicity in solving the radiative transfer problem,
compared with the direct numerical integration of the basic radiative transfer 
equation. However, one severe limitation of the technique is
that a large amount of computing time is required for the cases 
with extremely high optical depths. 
In order to handle the Ly$\alpha$ transfer for
extremely large $\tau_0$ using currently available computers,
we modified the prototypical Monte Carlo code 
developed in our previous papers \citep{all00,all01}.

\notetoeditor{Figure 1 should appear here}

\subsection{Acceleration Scheme for Extremely Thick Medium}

As was shown numerically in the previous papers \citep{all00,all01}
and explained in accordance with \citet{ada72}, in extremely 
thick media, Ly$\alpha$ photons escape from the media
by both frequency diffusion and spatial diffusion.
When a Ly$\alpha$ photon is emitted with the line
center frequency in an optically thick medium,
then it experiences a large number of core scatterings.
During the core scatterings, the photon does not move much
because the line-center opacity is very large.
During a series of a large number of core scatterings,
the Ly$\alpha$ photon can be scattered by a rapidly moving
atom in the tail of the Maxwell-Boltzmann velocity distribution.
Once the photon is scattered by such an atom, its frequency
changes abruptly and it becomes a wing photon.
For a wing photon, the hydrogen medium becomes more transparent
and the photon traverses a longer distance. 
In a moderately thick medium with $10^3 < \tau_0 < 10^3/a$,
where $a$ is the Voigt parameter, 
Ly$\alpha$ photons escape from the medium during at most a few wing 
scatterings; but in extremely thick media with $a\tau_0 > 10^3$,
the wing photon can not escape and have to experience 
a number of continuous wing scatterings, whose number is 
approximately given by $\langle N \rangle =x_s^2$ 
assuming random walk approximation \citep{ada72}.
Here $x_s=(a\tau_0)^{1/3}$ is the frequency at the peaks of profiles.

During such wing scatterings some photons become core
photons again by the so called `restoring force' \citep{ada72},
and then these photons suffer a large number of successive core scatterings.
One cycle of these processes is called `an excursion' \citep{ada72}, and
the photon spatially transfers by the repetition of excursions before 
it eventually achieves its escape from the medium during the wing 
scatterings: this final step of escape is called `single longest excursion'.

Considering these transfer processes, we can see that 
the major consumption of time in the Monte Carlo code occurs in the step 
of the computation of the outgoing wavevector associated with resonant or 
core scatterings, whose number of successive core
scattering is order of $\tau_0^2$. Therefore, the computing time 
can be significantly saved if we simplify this step in a clever way.
Our code was developed to distinguish the resonant core scattering
from the non-resonant wing scattering. Thus, in order to save
the computing time, whenever the photon experiences a resonant scattering
during its transfer, we can skip following successive 
resonant scatterings and to describe the next scattering 
by a fast moving hydrogen atom.

We now explain the skipping procedure in detail. 
The main idea for the procedure 
is to assume that any photons with $|x_1| < x_c$ suffer a large 
number of core scatterings locally and finally emerges 
from the local scattering region with frequency shift $|x_2|>x_c$
after a scattering with a rapidly moving atom, where $x_c$ is the critical
frequency shift which defines the borderline between core and wing
photons. In the accelerating scheme, we just skip the core scatterings 
and assigns $x_2$ in an appropriate way.
For the Lorentzian profile of Ly$\alpha$ scattering, 
$x_c$ can be safely set to $0 < x_c \le \sqrt{\pi}$ 
for moderately thick media with $10^3<\tau_0<10^3/a$, 
and to $x_c = \sqrt{\pi}$ for extremely thick media with $a\tau_0>10^3$.
Here we obtain $\sqrt{\pi}$ from the Eq.~(10) by substituting 
the effective wing optical depth $\tau_w=1$.
We find that the slightly larger $x_c$ can be 
permitted for even larger $\tau_0$. 

In order to describe the scattering by a fast moving atom, we select 
the velocity ($v$) of the scattering atom from the volcano-type 
thermal velocity distribution \citep{ave68}
\begin{equation}
P(v) = {1 \over \sqrt{\pi}} \exp(-v^2),
\end{equation}
where $x_c \le v \le v_{max}$, rather than from a simple Gaussian 
distribution. Here $v_{max}$ is safely set to be as large as $v_{max}=10$.
We call this method `the accelerating scheme'.
We note that this method is valid for $a\tau_0\ll 10^3$
only when $x_c \rightarrow 0$, which is tantamount to restoring 
the original code.

\subsection{Angular Redistribution and Polarization}

In this subsection, we provide the details of the angular redistribution
and polarization of the scattered radiation.
Our code treats the scattering of Ly$\alpha$ photons in a manner
very faithful to the atomic physics associated 
with the fine structure of hydrogen.
The level splitting of $^2P_{{1\over 2},{3\over 2}}$,
the excited state of Ly$\alpha$ transition, is $10{\rm GHz}$,
which amounts to the Doppler width of $1.3{\rm\ km\ s^{-1}}$.
Because this is much smaller than the thermal speed 
in a medium with $T\ge 100{\rm\ K}$,
one normally neglect the $^2P_{{1\over 2},{3\over 2}}$ level splitting,
treating it as a single level. However, even though $T < 100{\rm\ K}$,
the level splitting can be negligible when the medium has a large
optical thickness.

As is described in \citet{all00,all01},
in our Monte Carlo code we compute the probability
that the given photon is resonantly scattered with one of the two
levels $P_{{1\over 2},{3\over 2}}$ and the probability 
that the scattering occurs in the damping wings
at each scattering event.
We adopt the density matrix formalism to describe the angular distribution
and polarization of the scattered Ly$\alpha$, where the density operator
is represented by a $2\times2$ Hermitian matrix.
The density matrix element associated with the scattered photon
$\rho_{\beta\beta'}$ is related with that of the incident photon
$\rho_{\alpha\alpha'}$ by
\begin{equation}
\rho_{\beta\beta'}\propto\sum_{I,I'=P_{3/2},P_{1/2},\alpha,\alpha'}
{(\hat{\bf r}\cdot\epsilon^{(\beta')})_{AI}(\hat {\bf r}\cdot
\epsilon^{(\alpha)})_{IA}
\over E_I-E_A-\hbar\omega}
\rho_{\alpha\alpha'}
{(\hat{\bf r}\cdot\epsilon^{(\alpha')})_{I'A}^*(\hat{\bf r}\cdot
\epsilon^{(\beta)})_{AI'}^*
\over E_{I'}-E_A-\hbar{\omega}}.
\end{equation}
Here, the matrix elements due to other $nP (n>2)$ terms are neglected, and
the radial part of the matrix elements $<2p \| \  r\  \| 1s>$ is treated
as a constant. The relevant Clebsch-Gordan coefficients are found in
\citet{con51,lee94}.

As is shown by \citet{ste80}, scattering of Ly$\alpha$ in the damping wings
is characterized by the classical Rayleigh scattering 
phase function due to the
quantum interference between the two levels $P_{{1\over2}, {3\over2}}$.
In this case, one can show that the Ly$\alpha$ scattering 
in the wings can be regarded
as a resonance scattering between two levels with $J=0$ and $J=1$ as far as
the scattering phase function is concerned. The computation of the 
matrix elements
in this case is straightforward and the result is
\begin{eqnarray}
\rho'_{11} &=& \cos^2\Delta\phi \  \rho_{11}
-\cos\theta\sin2\Delta\phi \  \rho_{12} +
\cos^2\theta\sin^2\Delta\phi \  \rho_{22}  \nonumber \\
\rho'_{12} &=&{1\over 2}\cos\theta'\sin2\Delta\phi \  \rho_{11}
+(\cos\theta\cos\theta'\cos2\Delta\phi +\sin\theta\sin\theta'\cos\Delta\phi)
\  \rho_{12} \nonumber \\
&& -\cos\theta(\sin\theta\sin\theta'\sin\Delta\phi+
{1\over2}\cos\theta\cos\theta'\sin2\Delta\phi)\  \rho_{22} \\
\rho'_{22} &=& \cos^2\theta' \sin^2\Delta\phi \  \rho_{11}
+ \cos\theta'(2 \sin\theta\sin\theta'\sin\Delta\phi +
\cos\theta\cos\theta'\sin2\Delta\phi) \  \rho_{12} \nonumber  \\
&& +(\cos\theta\cos\theta'\cos\Delta\phi + \sin\theta\sin\theta')^2
\  \rho_{22} \nonumber,
\end{eqnarray}
where the incident radiation is characterized by the wavevector
$\hat{\bf k}_i =(\sin\theta\cos\phi, \sin\theta\sin\phi, \cos\theta)$
and the outgoing wavevector $\hat{\bf k}_f$ is correspondingly given with
angles $\theta'$ and $\phi'$ with $\Delta\phi = \phi'-\phi$.
Here, we assume that the circular polarization is always zero due to
the azimuthal symmetry in the scattering geometry and initial condition.

In the case of resonance scattering between $J={1\over2}$
and $J={3\over2}$ levels,
a straightforward calculation shows that the density matrix elements
associated with the outgoing photons are given by
\begin{eqnarray}
\rho'_{11} &=& (5+3\cos2\Delta\phi)\rho_{11} \nonumber \\
&&   +[(5-3\cos2\Delta\phi)\cos^2\theta+2\sin^2\theta] \rho_{22}
        -6\cos\theta\sin2\Delta\phi \rho_{12} \nonumber \\
\rho'_{12} &=& 3\sin2\Delta\phi\cos\theta' \rho_{11}
          +6(\cos\theta\cos\theta'\cos2\Delta\phi+\sin\theta\sin\theta'
          \cos\Delta\phi) \rho_{12} \nonumber \\
&&        +3\cos\theta(-2\sin\theta\sin\theta'\sin\Delta\phi-
          \cos\theta\cos\theta'\sin2\Delta\phi) \rho_{22} \nonumber \\
\rho'_{22} &=& [(5-3\cos2\Delta\phi)\cos^2\theta'+2\sin^2\theta') \rho_{11}
             \nonumber \\
&&            +[(5+3\cos2\Delta\phi)\cos^2\theta\cos^2\theta'+
              2\cos^2\theta\sin^2\theta' \nonumber \\
&&            +12\cos\Delta\phi \cos\theta'
              \cos\theta\sin\theta\sin\theta'+2\cos^2\theta'\sin^2\theta
              +8\sin^2\theta\sin^2\theta'] \rho_{22} \nonumber\\
&&            +(6\sin2\Delta\phi\cos\theta\cos^2\theta'+
              2\sin\Delta\phi\cos\theta'\sin\theta\sin\theta')\rho_{12}.
\end{eqnarray}

When the scattering is resonant with the level $P_{1\over 2}$, then the
scattered radiation is characterized with the isotropic angular distribution
and hence completely unpolarized. Therefore, the density matrix is constant
and ${1\over2}$ times the identiy matrix irrespective of 
the incident and outgoing wavevectors.

The density matrix elements associated with the outgoing photon given
in Eqs.~(3-5) are normalized by the unit trace condition.
It is readily verified
that a $90^\circ$ scattering of a completely unpolarized incident
photon yields 100 percent polarization in the case of a wing scattering
and the degree of polarization in the case of a resonant scattering
between $S_{1\over 2}\rightarrow P_{3\over 2}$ becomes ${3\over7}$
(e.g. \citet{cha60,lee94}).

\notetoeditor{Figure 2 should appear here.}

In Figure~2 we compare the results obtained by using 
the accelerated code with those calculated by using the original 
code in our previous works. Here we set the temperature of the
scattering media to be $T=10{\rm\ K}$ or $a=1.47\times10^{-2}$,
and the photon source is located at the origin of the central
plane of the slab, which is the same configuration to the midplane 
sources by symmetry.
In the figure the solid lines stand for those results
without the accelerating scheme, and the dashed lines represent
the results using the modified code with $x_c=1$.
Notable deviation between the two results is seen 
for $a\tau_0\le 1.47\times10^2$ or $a\tilde\tau_0\le 2.6\times10^2$.
Here $\tilde\tau_0$ is the line-center optical depth defined 
in terms of the normalized Voigt function 
in the previous researches \citep{ada72,neu90}, 
where $\tilde\tau_0=\sqrt{\pi}\tau_0$.

The original code can be regarded as a special case of 
the accelerated code when $x_c\rightarrow 0$.
In order to check this, we calculate the emergent profile for the 
smaller core-wing boundary frequency, $x_c$.
For the case with $\tau_0=10^3$, we show 
our result with $x_c=0.6$ by the dotted line in Figure~2. 
Note that the dashed line for $\tau_0=10^3$ is obtained for $x_c=1$.
In the figure we confirm that the emergent profile converges 
to that from the original code as $x_c\rightarrow 0$. 
It is clear that the accelerated code is much more efficient
than the original code, and it gives the correct results.

\section{Line Transfer in Extremely Thick Media}

\subsection{Emergent Profiles}

In this subsection, we compare our results with analytic solutions 
and check the validity of our accelerating scheme.
In extremely thick media with $a\tau_0 > 10^3$, wing scatterings dominate
the line transfer in space and the Voigt function $H(x,a)$ can be 
approximated by
\begin{equation}
H(x,a) \simeq { a \over \sqrt{\pi} x^2},
\end{equation}
where the optical depth at a frequency shift $x$ is given 
by $\tau_x=\tau_0 H(a,x)$.
\citet{ada72,har73,neu90} introduced the diffusion
approximation according to which the Ly$\alpha$ photons transfer in space
only by wing scatterings, and the distance traversed per scattering
is described by Eq.~(6). \citet{neu90} derived an analytic solution 
for the extremely thick media,
\begin{eqnarray}
J(\pm\tau_0,x) = {\sqrt{6} \over 24}{x^2 \over a\tau_0}
{ 1 \over \cosh[(\pi^4/54)^{1/2}(|x^3|/a\tau_0)]}.
\end{eqnarray}

We perform Monte Carlo calculations for a monochromatic source
with its input frequency $x_0=0$.
We simulate several cases with different $a\tau_0$ values
so that both moderately thick and extremely thick cases are included
in the computations. In each case, the source is located 
at the midplane of a static neutral slab.

First, we set the temperature of the scattering medium to be $T=10 {\rm\ K}$,
which corresponds to $a=1.49\times10^{-2}$.
The relevant situation for this case may be found in the later evolutionary 
epoch of starburst galaxies,
when the cold supershell with large H~I column density expands
very slowly. The column density ($N_{\rm HI}$) of normal galaxies
and dwarf galaxies ranges $N_{\rm HI}=10^{19}-10^{22} \cm^{-2}$, which corresponds to
the Ly$\alpha$ line-center optical depth $\tau_0=10^6-10^9$.
The line-center optical depth is related to the H~I column density
$N_{\rm HI}$ via
\begin{equation}
\tau_0 \equiv 1.41\ T_{4}^{-1/2}
\left[{N_{\rm HI} \over {10^{13} \rm cm^{-2}}}\right],
\end{equation}
where $T_4 = T/10^4 {\rm K}$ and $T$ is the temperature of the scattering
medium.

\notetoeditor{Figures 3a and 3b should appear here.}

Here we choose $\tau_0=10^3, 10^4, 10^5, 10^6, 10^7$
or $a\tau_0=1.49\times10$, $1.49\times10^2$, $1.49\times 10^3$,
$1.49\times 10^4$, $1.49\times 10^5$.
We note that these are transformed respectively
into $a\tilde\tau_0=2.64\times10,\ 2.64\times10^2$, $2.64\times10^3$,
and $2.64\times10^4$.

Our results are shown in Figure~3a and Figure~3b.
We see that the profiles of our Monte Carlo results are in good agreement
with the analytic solutions for $a\tau_0 \ge 10^3$.
For $\tau_0=10^3$, we also show the emergent profile for the case
of $x_c=0.6$ which is closer to the result of original code 
in \citet{all00,all01}.
We see that it also shows deviation from the analytic solution
derived by \citet{neu90}, which is expected due to the invalidity 
of diffusion approximation for $a\tau_0 <10^3$.
We conclude that our accelerating scheme is valid 
for the various range of optical depths.

\notetoeditor{Figure 4 should appear here.}

Next, we perform Monte Carlo calculations for hotter media,
whose Voigt parameter is given by $a=4.71\times10^{-4}$
corresponding to $T=10^4{\rm\ K}$.
We choose $\tau_0=10^6, 10^7, 10^8$, and $10^9$, which correspond
to $a\tau_0=4.71\times10^2, 4.71\times10^3, 4.71\times10^4$,
and $4.71\times10^5$. We note that these are transformed respectively
into $a\tilde\tau_0=8.35\times10^2,\ 8.35\times10^3$, $8.35\times10^4$,
and $8.35\times10^5$.
In Figure~4, we show the results of our accelerated
Monte Carlo code and compare them with the analytic solution, Eq.~(7).
The solid lines represent the results of our Monte Carlo calculations,
and the dotted lines do analytic solutions.
We see that our numerical results are in very good agreement 
with analytic solutions for very thick media ($\tau_0\ge 10^8$).
In extremely thick media, we can safely assume that only wing scatterings 
contribute to the spatial transfer of Ly$\alpha$ photons,
and so our accelerated code can describe
the line transfer process very well.

\notetoeditor{Figure 5 should appear here.}

Third, we also perform calculations for sources uniformly distributed 
along the vertical direction of the slab.
The relevant application for this case can be found in 
the giant H~II region,
where the Ly$\alpha$ sources and the scattering media
may be mixed together. We assume that the typical temperature
of H~II region is $T=10^4 {\rm\ K}$, and that the size of
H~II region is ${\rm kpc}$ scale. In this environment 
the degree of ionization can lead to the volume density of
neutral hydrogen $n_{\rm HI}=1\cm^{-3}$. Hence the total column density 
of the H~II region can be $N_{\rm HI}\sim10^{20} \cm^{-2}$,
which corresponds to the line-center optical depth 
of $\tau_0 \sim 10^4-10^7$.
In the code, we generate the sources at the location $(p_x,p_y,p_z)=
(0,0,\tau_s)$, where $\tau_s=(2R-1)\tau_0$ and $R$ is the uniform random
number having range of $[0,1]$.

When photon sources are located near the surface, the optical depth 
toward the nearest boundary is much less than that toward the midplane 
of the slab. Hence, we adjust the core-wing boundary frequency ($x_c$)
in accordance with the vertical location of photon sources 
which is denoted by $p_z$:
we used $x_c = A(1-p_z/\tau_0)$, where $A$ is a core-wing boundary
frequency in units of thermal velocity for midplane sources. This formula
gives correct behaviors for both thin and thick cases.
Clearly $x_c\rightarrow 0$ as $|p_z|\rightarrow\tau_0$.

In Figure~5 we show our results only for the extremely
thick cases with $\tau_0=10^7, 10^8$ and $a=4.71\times10^{-4}$.
We note that the results for the moderately thick cases can
be found in the previous paper \citep{all01}.
Our results (solid lines)
are compared with the analytic solutions (dashed lines).
From the analytic solutions for the case of midplane source \citep{neu90},
we obtain the analytic solutions for the case of distributed sources
by convolving the midplane solution, Eq.~(7), with the uniform source 
positions ($\tau_s$),
\begin{equation}
J_d(\pm\tau_0,x)=\int^{\tau_0}_{-\tau_0}
J_m (\pm\tau_0,x;\tau_s) \delta(\tau_s) d\tau_s.
\end{equation}

\notetoeditor{Figure 6 should appear here.}

In the figure we can see that our numerical results agree very well
with the analytic ones, and that the peaks are more shifted to the line-center 
compared to the midplane cases.
This is because the sources near the surface contribute to
the central part of emergent profiles.

\citet{ada72} argued that in extremely thick media with $a\tau_0>10^3$,
the emergent flux depends on the product $a\tau_0$, not on
the individual parameters. This is evidently seen in Eq.~(7), 
in which $J(\pm\tau_0,x)$ is dependent upon $a\tau_0$.
In order to check this fact,
we perform Monte Carlo calculations for a few pairs of $a$ and $\tau_0$
with common $a\tau_0$ values.
For $a\tau_0=4.71\times10^2$ (or $a\tilde\tau_0=8.35\times10^2$),
we perform calculations for two pairs;
($a=4.71\times10^{-4}$, $\tau_0=10^6$) and
($a=4.71\times10^{-2}$, $\tau_0=10^4$).
For $a\tau_0=4.71\times10^3$ (or $a\tilde\tau_0=8.35\times10^2$),
we choose ($a=4.71\times10^{-4}$, $\tau_0=10^7$)
and ($a=4.71\times10^{-2}$, $\tau_0=10^5$).
The other pair is ($a=4.71\times10^{-4}$, $\tau_0=10^8$) and
($a=4.71\times10^{-2}$, $\tau_0=10^6$), which have $a\tau_0=4.71\times10^4$.
Our results are shown in Figure~6, where we see that,
for the extremely thick cases $a\tau_0\ge10^3$, the line profiles 
for each pair with the same $a\tau_0$ agree very well with each other.
Thus we confirm that our code reproduces the results in \citet{ada72} 
very well.

\notetoeditor{Figure 7 should appear here.}

Finally, we also consider the radiative transfer of continuum photons
near Ly$\alpha$. We simply assume that an input spectrum is flat.
In Figure 7 we show our results for $\tau_0=10^4, 10^5, 10^6, 10^7$
fixing $a=1.49\times10^{-2}$, where symmetric double peaks and 
one absorption trough at the line-center are seen.
The absorption trough broadens as $\tau_0$ gets larger.
We show the wavelength ($\lambda$) in units of ${\rm \AA}$ 
on the top edge of the figure,
which is scaled by $\lambda \propto T^{1/2}$.
Here $T$ is the temperature of the scattering medium. Note again 
that the spectra are symmetric relative to the line-center.

\subsection{Limb brightening and darkening}

\notetoeditor{Figure 8 should appear here.}

In this subsection we consider the directionality of emergent 
Ly$\alpha$ photons. In the Monte Carlo code we collect 
emerging photons with the directional information 
into the equally spaced bins in $\mu\equiv\cos\theta$
where $\theta$ is the angle between the normal direction of the
slab and the wavevector of the photons. Since the number of photons
along $\mu$ is proportional to $I(\mu) \cos\theta d\Omega$,
where $\Omega$ is the solid angle, we 
obtain the intensity $I(\mu)$ by dividing the
number of photons by $\mu$. We call the function
$I(\mu)$ the `directionality'. 
This definition of directionality is different from that used by 
\citet[hereafter PM86]{phi86},
who defined it as the intensity 
normalized with respect to the intensity normal to the slab.

The results are shown in Figure~8.
The curves for $a\tau_0\le 10^3$ are nearly isotropic, because most Ly$\alpha$
photons escape from the medium by single longest flights: each
Ly$\alpha$ photon is scattered locally near the source and becames
a wing photon by frequency diffusion, when the medium gets transparent.
Therefore, a large number of local scatterings isotropize the radiation field,
which is characterized by no or (at least a negligible) polarization.

As $\tau_0 >10^3$, the directionality gradually converges to that of the
Thomson-scattered radiation emergent from a Thomson-thick electron medium
represented by the thick solid line. This Thomson limit was obtained by
\citet{cha60} and was confirmed numerically by PM86. PM86 also provided
the results for optically thin as well as optically thick cases.
Therefore, we compare our results with those from the radiative
transfer in a pure electron scattering medium, because Thomson scattering
shares the same Rayleigh phase function as Ly$\alpha$ wing scattering.

It is convenient to consider 
the effective wing optical depth, $\tau_w$, which is the optical
depth at the effective frequency of emergent photons, $x_s=(a\tau_0)^{1/3}$.
From Eq.~(6) we can see that 
\begin{equation}
\tau_w = {1 \over \sqrt\pi}(a\tau_0)^{1/3}={1 \over \sqrt\pi}x_s.
\end{equation}
We compare this optical depth with the optical depth 
for the Thomson scattering defined by 
$\tau_e = \sigma_T N_e$,
where $\sigma_T= 0.665\times10^{-24} {\rm\ cm^2}$  is
the Thomson scattering cross section,
and $N_e$ is the electron column density.

When $a\tau_0$ is small, after a large number of local resonant scatterings,
a Ly$\alpha$ photon becomes a wing photons due to the scattering with a fast 
moving hydrogen atom. Now, for this wing photon the slab is a transparent
medium, especially along the normal direction and escape to this direction
occurs without subsequent scatterings. However, to the grazing direction,
because of simple geometrical considerations, there can be a significant
probability of further wing scatterings before escape. Therefore,
the situation becomes quite similar to the electron scattering in a 
plane-parallel slab that is quite thin and illuminated from the other side. 
Limb brightening is a natural consequence because we see more sources 
in the grazing direction, as is shown in Figure~8.

For the extremely thick cases with $a\tau_0>10^3$, the effective wing optical 
depth is estimated to be $\tau_w>6$ by Eq.~(10). This corresponds 
to the result of the Thomson scattered radiation in an electron cloud 
with its optical depth $\tau_e>6$. This limiting behavior of 
the directionality may be summarized as
\begin{equation}
{I(\mu)\over I(\mu=1)} \simeq {1\over3}(1+2\mu),
\end{equation}
which is an approximate relation proposed by PM86 for
the numerical solution introduced in the Table XXIV of \citet{cha60}.
This is shown by a thick solid line in Figure~8.
We see that the directionality of Ly$\alpha$, being transferred 
in an extremely thick hydrogen medium, converges to that 
for the case of optically thick electron cloud. In this optical depth regime,
the optical depth for Ly$\alpha$ wing photons becomes large, 
and the photons spatially transfer in the medium by random walks.
Hence, the escape of photons occurs preferentially to the normal to the slab,
where the opacity is relatively small. We regard this phenomenon 
as a limb darkening, which was also called a `beaming' by PM86.

\notetoeditor{Figure 9 should appear here.}

In an extremely thick medium, Ly$\alpha$ photons experience a number of
successive wing scatterings just before their escape from the medium
during their single longest excursion.
How many times do the photons experience the successive wing scatterings
just before escape? According to \citet{ada72}, in an extremely thick medium
with an isotropic Ly$\alpha$ source located at the midplane, photons 
acquire a typical frequency shift $x_s = (a\tau_0)^{1/3}$ during excursions
(see also section 2 of this paper).
In an extremely thick medium, the spatial transfer of Ly$\alpha$ photons
occurs during a series of wing scatterings, and so we can approximate
the spatial transfer by random walks. Thus from Eq.~(10) 
we can estimate the mean number of successive wing scatterings 
just before escape ($\langle N_w\rangle$) by
\begin{equation}
\langle N_w \rangle = \tau_w^2 = { 1 \over \pi} (a\tau_0)^{2/3}
\end{equation}
or
\begin{equation}
\log_{10}\langle N_w \rangle = {2 \over 3} \log_{10} (a\tau_0) - 0.497.
\end{equation}

\notetoeditor{Figure 10 should appear here.}

In the accelerated Monte Carlo calculations we count the number of wing
scatterings just before escape, and compare the result
with the above relationship.
In the upper panel of Figure~9 we show the spectral 
distribution of the number of wing scatterings just before escape.
We can see that the photons emerging with larger frequency shift have
been experienced more wing scatterings just before escape
than those with smaller frequency shift.

In order to check the relationship of Eq.~(13), we show in Figure~10
the graph of $\log_{10} \langle N_w \rangle$ vs. $\log_{10} a\tau_0$,
where $\langle N_w \rangle$ is the value averaged over frequency,
and $a\tau_0$ represents the importance of wing scatterings 
in Ly$\alpha$ line transfer. Here the number of photons for each point is 4000,
and the error in $N_w$ is the flux weighted Poisson error.
The least square fitting gives us the following relationship,
\begin{equation}
\log_{10}\langle N_w \rangle = (0.674\pm 0.010) 
\log_{10} a\tau_0 - (0.552\pm 0.042).
\end{equation}
We see that the power and the coefficient in Eqs.~(13) and Eq.~(14)
are in good agreement with each other.

\section{Discussion}

We have investigated the Ly$\alpha$ line transfer in extremely thick,
dustless, uniform, and static media. 
In order to reduce the calculation speed, we developed an accelerating
scheme for our previous Monte Carlo code, in which we skip
an extremely large number of core scatterings that may occur during the line 
transfer. The code also deals with the angular redistribution accurately 
by using the density matrix formalism
where the phase functions calculated in a manner faithful to the 
atomic physics. The modified code passes the requirement in both
efficiency and accuracy when calculating the Ly$\alpha$ line profiles 
emergent from extremely thick media. We checked that the line profile
is dependent only upon $a\tau_0$, which was argued by \citet{ada72}.
With the code we obtained solutions for both the midplane source and
the uniformly distributed sources, both of which show excellent
agreement with the analytic solutions derived by \citet{neu90}.
In the latter case, the peaks tend to shift toward the line-center 
due to the sources located near the slab surface.
We also presented our results on the radiative transfer
of continuum photons near Ly$\alpha$, and the emergent spectra show
the double peaks and the broad absorption at the line-center.
However, it is natural to think that dust extinction will destroy 
a significant fraction of photons forming the double peaks,
and eventually a broad absorption trough will form.

We also examined the directionality of emergent Ly$\alpha$ photons.
When the medium is slightly thick with $a\tau_0 << 10^3$,
we observe a limb brightening effect, because the slab becomes transparent
to wing photons.
As $a\tau_0$ increase, the directionality eventually 
converges to that of the Thomson scattered radiation in a thick electron cloud
which was investigated by \citet{cha60}.
The limiting behavior of the curves is the beaming of Ly$\alpha$ photons
into the direction normal to the slab.
In short, the Ly$\alpha$ emission from the star-forming regions
that are surrounded by scattering neutral media shows the limb brightening
when $a\tau_0 \ll 10^3$, and the limb darkening when $a\tau_0 > 10^3$.
These facts can be physically explained, since both the Ly$\alpha$
scattering by neutral hydrogen at wings and the Thomson scattering 
in an electron cloud have the same Rayleigh phase function.

We also counted the mean number of successive wing scatterings 
just before escape, which is denoted by $\langle N_w \rangle$,
and by Monte Carlo calculations we confirmed the relationship 
$\langle N_w \rangle = {1 \over \pi} (a\tau_0)^{2/3}$,
which is derived from the diffusion approximation. 

In conclusion, it is important to consider
the quantum mechanical properties of Ly$\alpha$ scatterings and 
the exact transfer processes, and our accelerated Monte Carlo code
can afford those requirements as well as the computing time.

Our works can be applied to local and distant starburst galaxies.
Approximately half of their Ly$\alpha$ lines 
show broad absorption troughs in the Lorentzian wings.
These absorption troughs are believed to be formed by 
optically thick, static H~I media with some amount of dust.
However, the media in this study have been assumed to be dustless,
which is unrealistic for almost all situations.
Hence, we will take dust effects into account in the future works.
Moreover, the emergent profiles are also influenced by
the physical quantities such as porosity of dust and 
homogeneity of the medium as well as dust abundance and H~I column density,
so that we are now undertaking investigations about the effects of 
these components on the line formation.

Another half of the starburst galaxies, either nearby or remote,
show P-Cygni type Ly$\alpha$ emission.
As \citet{kun98} suggested, expanding media around
the star-forming regions play an important role in avoiding the destruction 
of Ly$\alpha$ photons by dust and forming the P-Cygni type Ly$\alpha$
profiles. When we calculated the Ly$\alpha$ line formation 
in an expanding medium, it is difficult to deal with the back-scattered 
photons. We have been investigating this topic by applying the formailsm
developed in this paper, and the results will be reported in another paper. 

In conclusion, Ly$\alpha$ from starburst galaxies bears
important and useful information on the physical environment 
of those systems including the kinematics, H~I
column density, abundance and porosity of dust  
in the surrounding media of the star-forming regions in those galaxies.
Being very luminous in addition to being sensitive to many physical quantities,
Ly$\alpha$ can be easily observed by using
the modern observational technology. 

\acknowledgments
This work was achieved as a part of doctorial dissertation
in the School of Earth and Environmental Sciences of Seoul National University
financially supported by Brain Korea 21 of the Korean Ministry of Education.
SHA and HML appreciate the support. Also this work is completed 
with the financial support of the Korea Institute for Advanced 
Study, which is pleasantly acknowledged.

\clearpage

\begin{figure}
\plotone{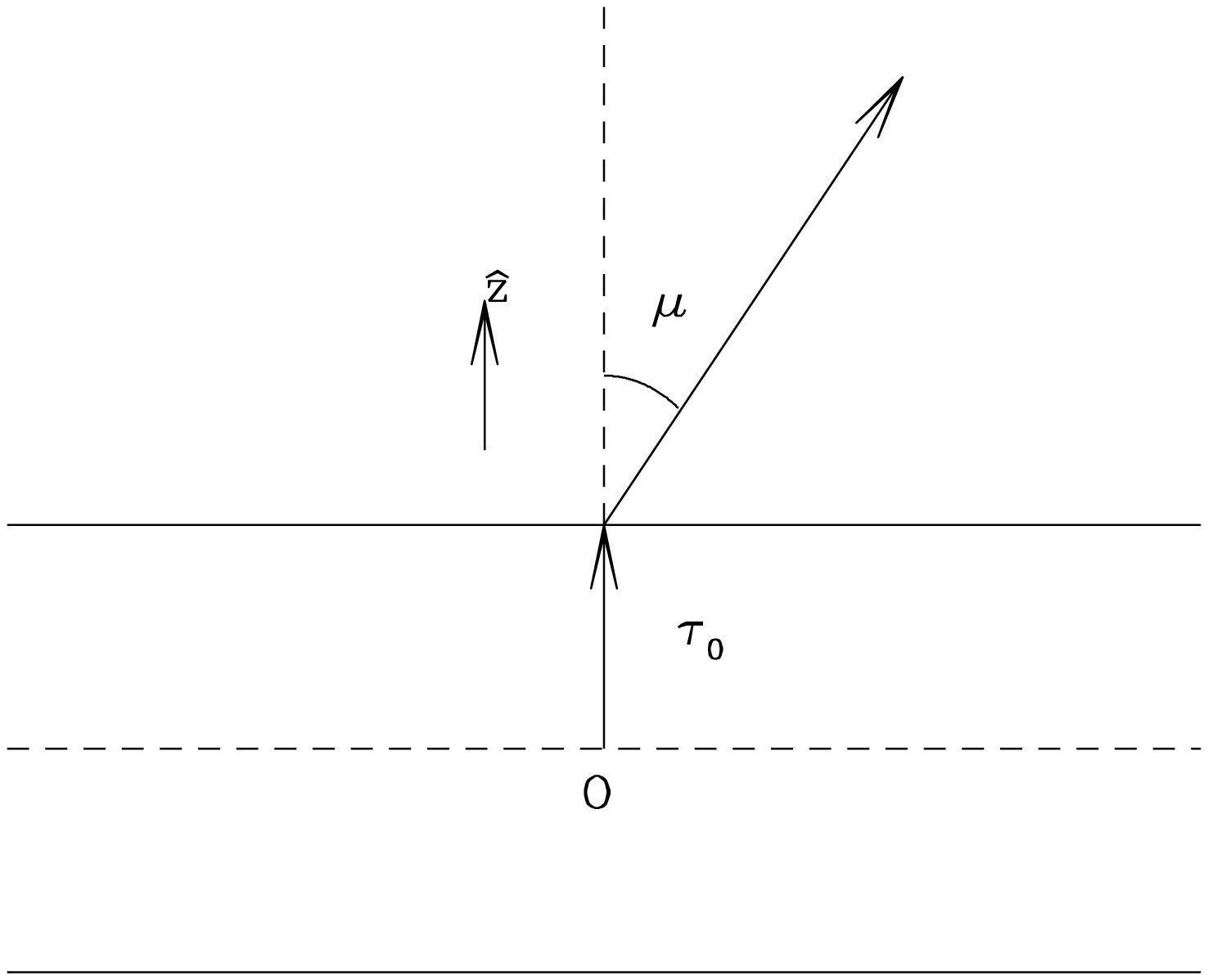}
\figcaption{Configuration studied in this work. We consider the static
and uniform slab. We also consider two types of source distribution:
in one type the Ly$\alpha$ source is located at the origin
denoted in the figure by $O$, and in the other they are uniformly scattered
along z-axis. The vertical thickness of
the slab is $2\tau_0$, and we define $\mu$ as the cosine
of the angle between the wave vector of the emergent photon
and the slab normal.
\label{fig1}}
\end{figure}

\clearpage

\begin{figure}
\plotone{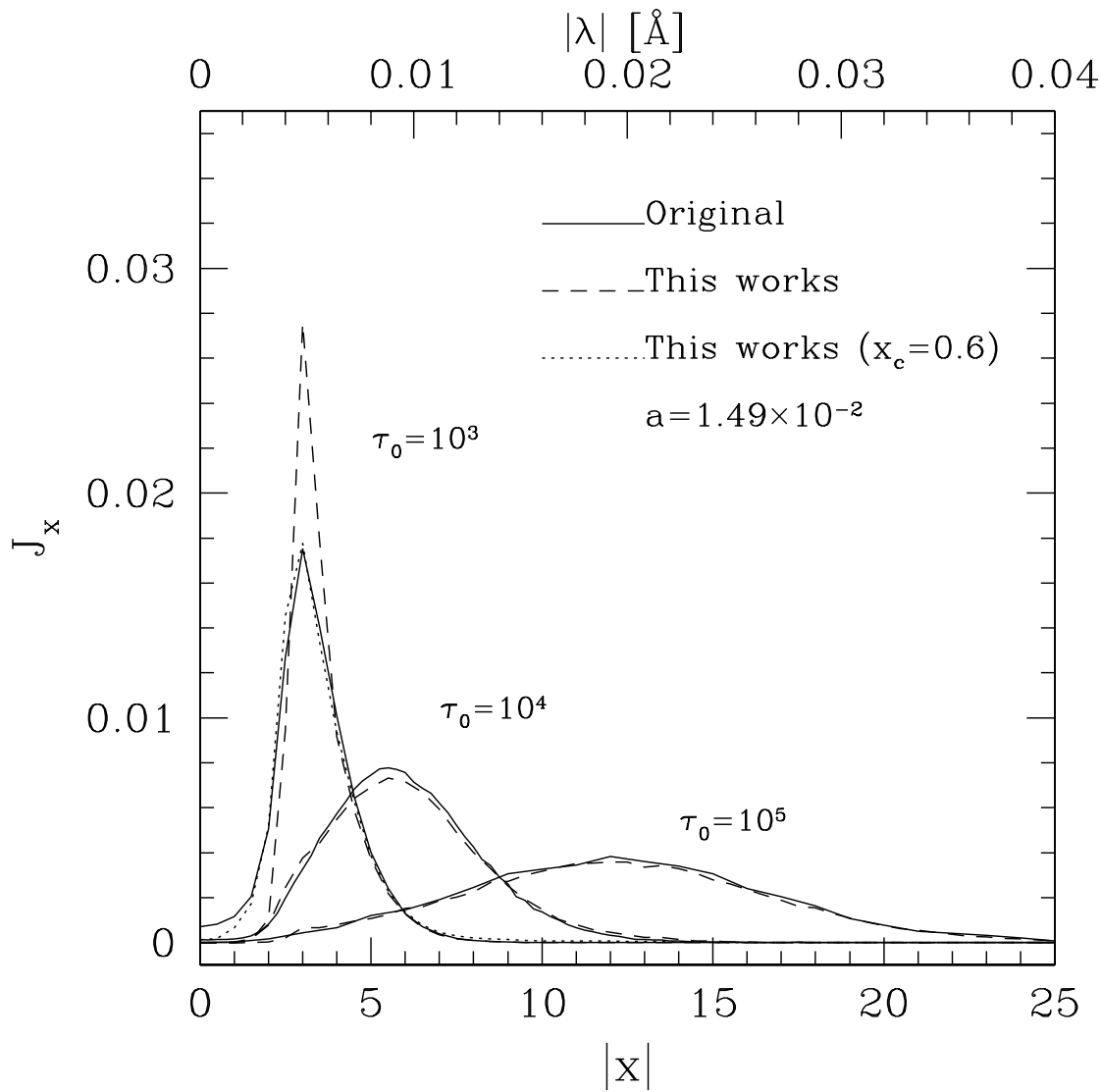}
\figcaption{
Comparisons between the results of the original code (solid line)
and those of the code with the accelerating scheme
(dashed line). The bottom horizontal axis represents frequency shift $x$
in units of the thermal width, and the top horizontal axis 
represents wavelength for $T=10 {\rm K}$.
Comparisons were made for the case of midplane
sources radiating monochromatic Ly$\alpha$ photons with $x_0=0$.
The dotted line represents the result for a small core-wing boundary
frequency, $x_c$.
\label{fig2}}
\end{figure}

\clearpage

\begin{figure}
\plotone{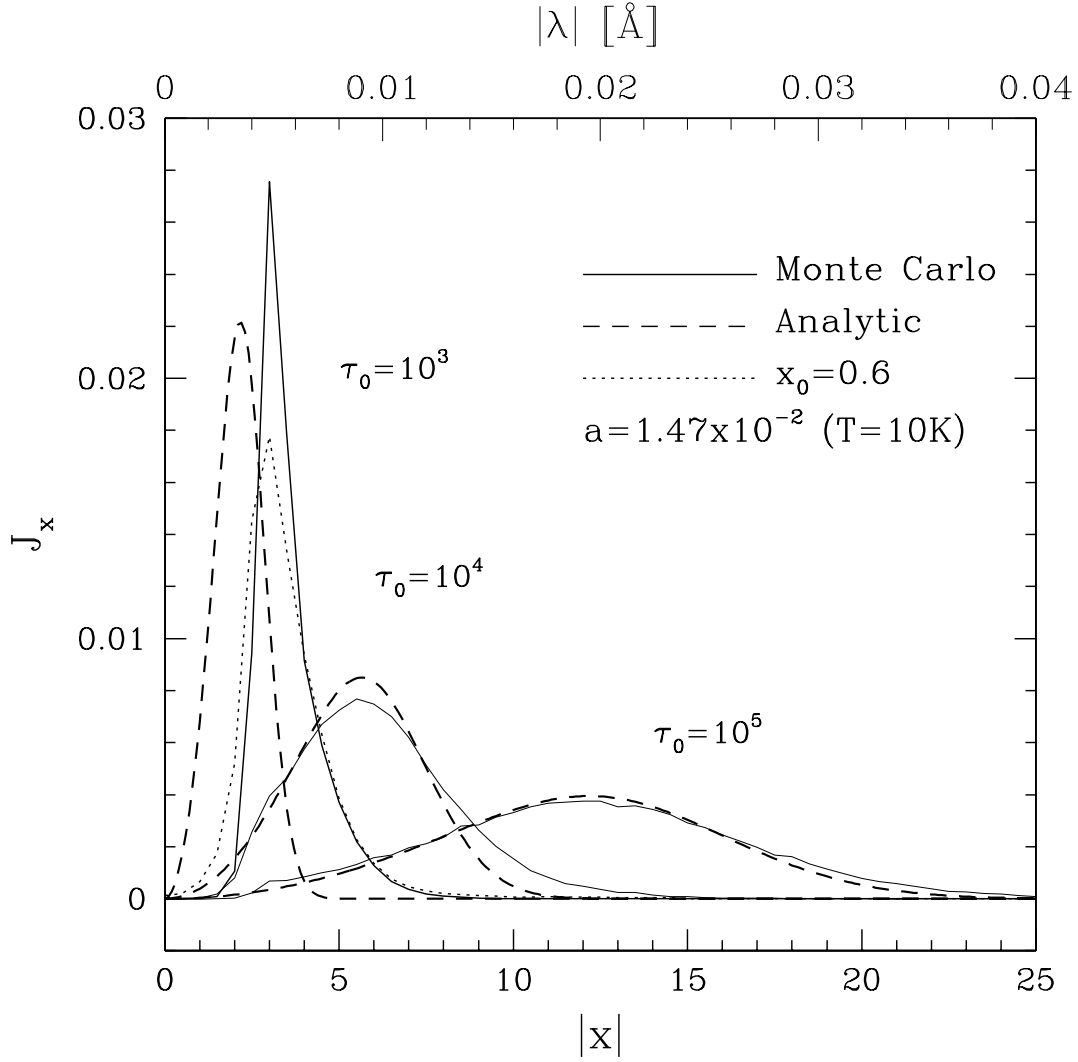}
\figcaption{
Ly$\alpha$ profiles emerging from thick media of different $\tau_0$'s,
where we fix $a=1.49\times10^{-2}$ or $T=10{\rm\ K}$.
The bottom horizontal axis represents frequency in units of the thermal width,
and the top horizontal axis represents wavelength shift for $T=10 {\rm K}$.
Here the solid lines stand for
results of our Monte Carlo calculations, and the dashed lines for
the analytic solutions given by Neufeld (1990).
The total flux of the line is normalized to $1/4\pi$ in accordance with
Neufeld's normalization. We note that the profiles are symmetric about
the origin, $x=0$.
\label{fig3a}}
\end{figure}

\clearpage

\begin{figure}
\plotone{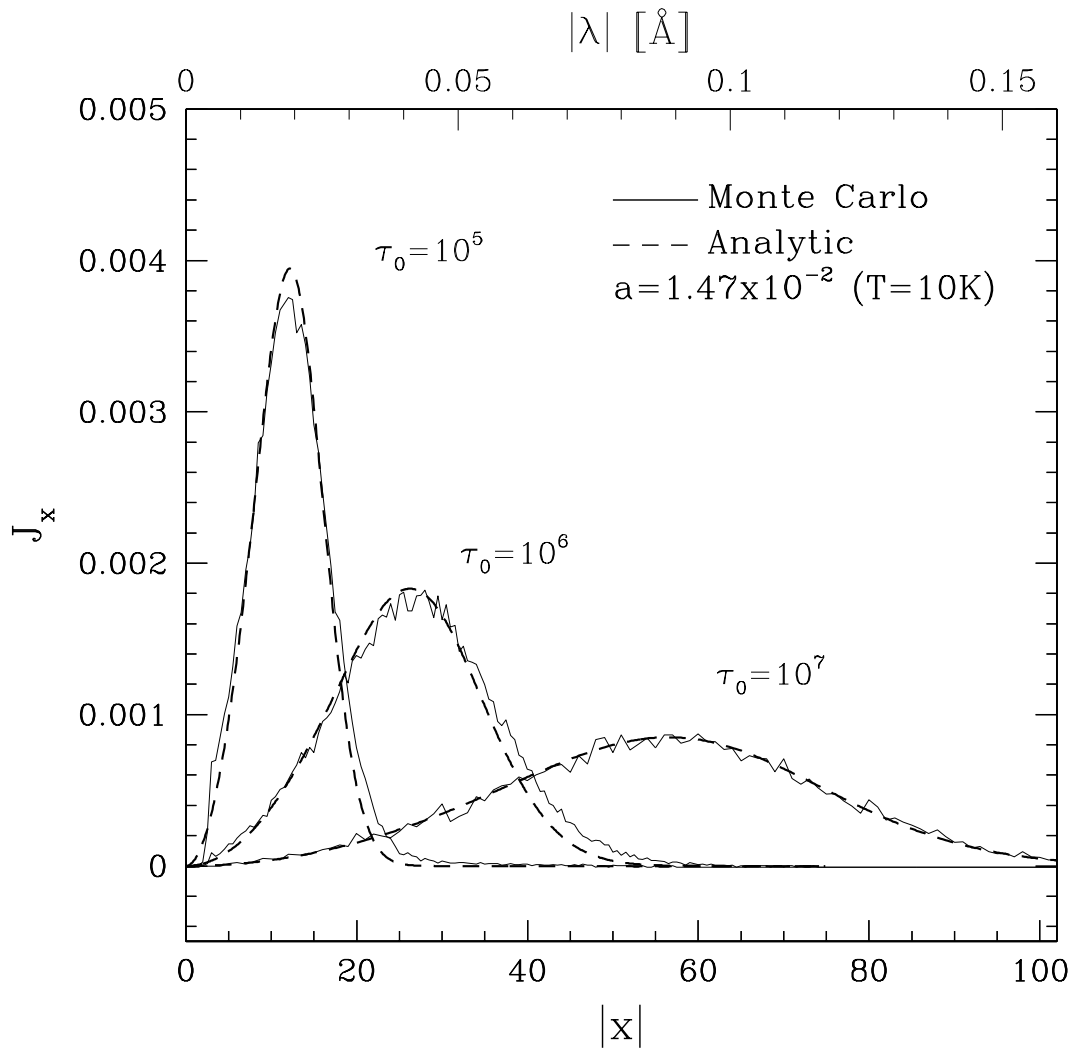}
\figcaption{Ly$\alpha$ profiles emerging from extremely thick media
with different $\tau_0$'s and a common Voigt parameter $a=1.49\times10^{-2}$
corresponding to $T=10{\rm\ K}$. Here the solid lines stand for
the results of our Monte Carlo calculations, and the dashed lines for
the analytic solutions given by Neufeld (1990).
Here horizontal axis represents frequency shift $x$ in units of the thermal 
Doppler width, and the total flux of the line is normalized to $1/4\pi$ 
in accordance with
Neufeld's normalization. Note that the profiles are symmetric about
the origin, $x=0$.
\label{fig3b}}
\end{figure}

\clearpage

\begin{figure}
\plotone{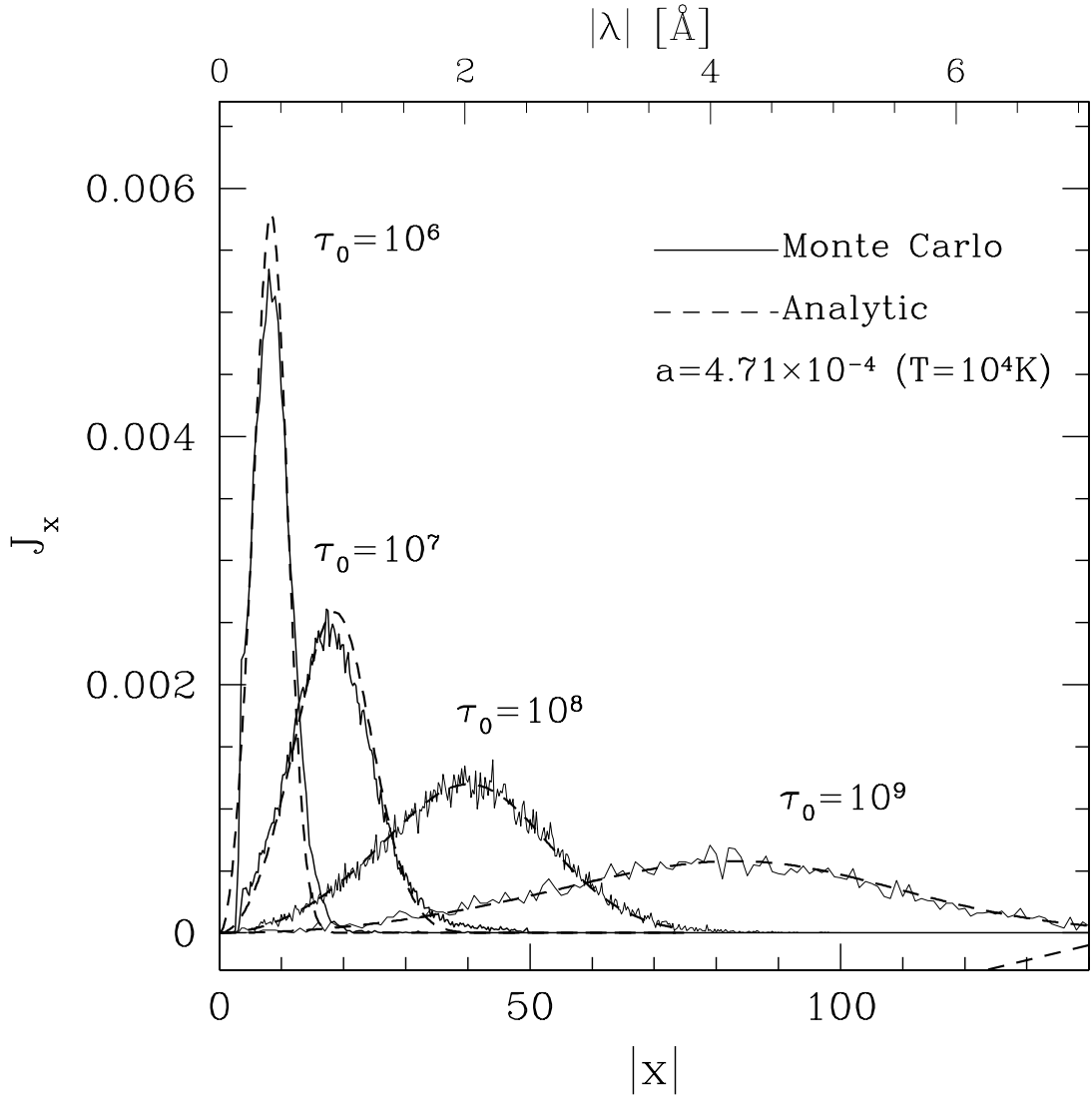}
\figcaption{Our emergent profiles (solid lines) are
compared with Neufeld's analytic solution (dotted lines).
\label{fig4}}
\end{figure}

\clearpage

\begin{figure}
\plotone{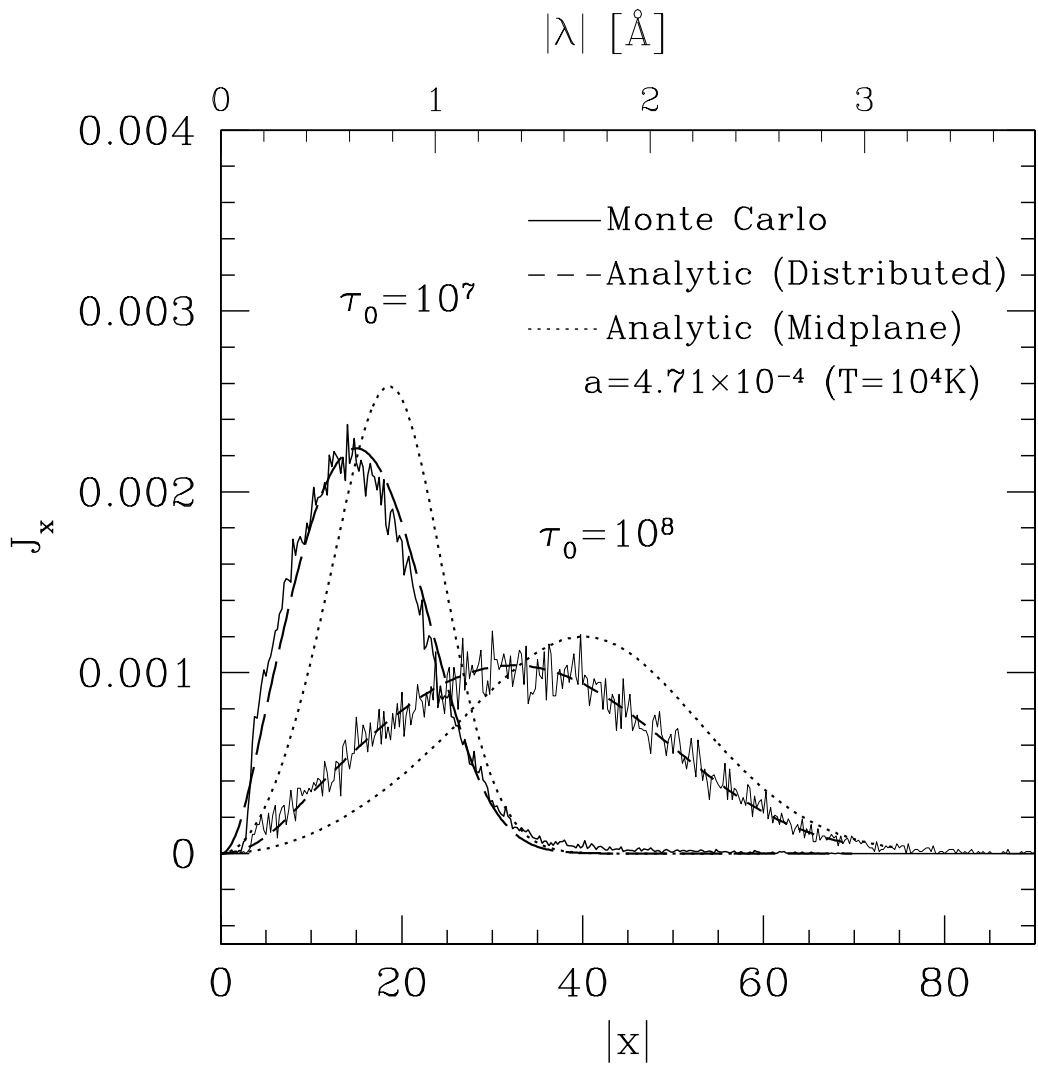}
\figcaption{
Emergent profiles for the cases of distributed sources.
Comparing with the midplane solutions of the same $a$ 
and $\tau_0$ (dotted lines), their peaks are shifted
to the line-center.
Here solid lines represent our Monte Carlo results,
and the dashed lines do the analytic solutions
obtained by convolving Neufeld's solution.
\label{fig5}}
\end{figure}

\clearpage

\begin{figure}
\plotone{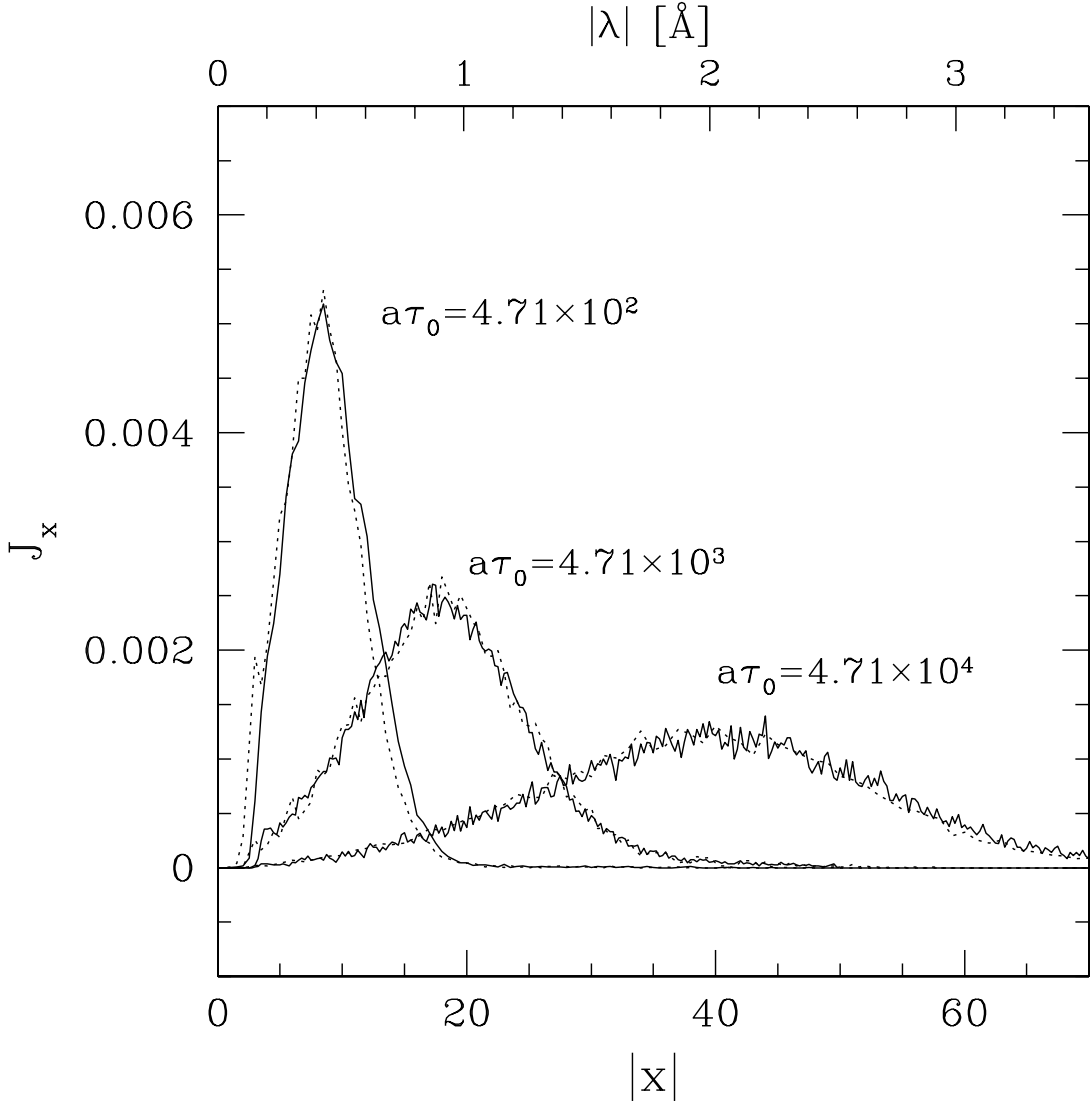}
\figcaption{
Emergent profiles of a few pairs of Monte Carlo simulations which have
the common $a\tau_0$ values. The pair of physical parameters in
the calculations are listed in the text. We see that the cases 
with the same $a\tau_0$ show the same emergent profiles.
Note that the profile is symmetric with respect to the origin $x=0$.
On the top horizontal axis is shown the wavelength ($\lambda$)
in units of ${\rm \AA}$, which scales $\lambda \propto T^{1/2}$.
Here $T$ is the temperature of the scattering medium,
and set to be $T=10^4 {\rm K}$ or $a=4.71\times10^{-4}$.
For $T=10 {\rm K}$ or $a=1.49\times10^{-2}$, the wavelengh
scales are smaller by a factor of $\sqrt{1000}=31.6$,
because $\Delta\nu_D\propto T^{1/2}$.}
\label{fig6}
\end{figure}

\clearpage

\begin{figure}
\plotone{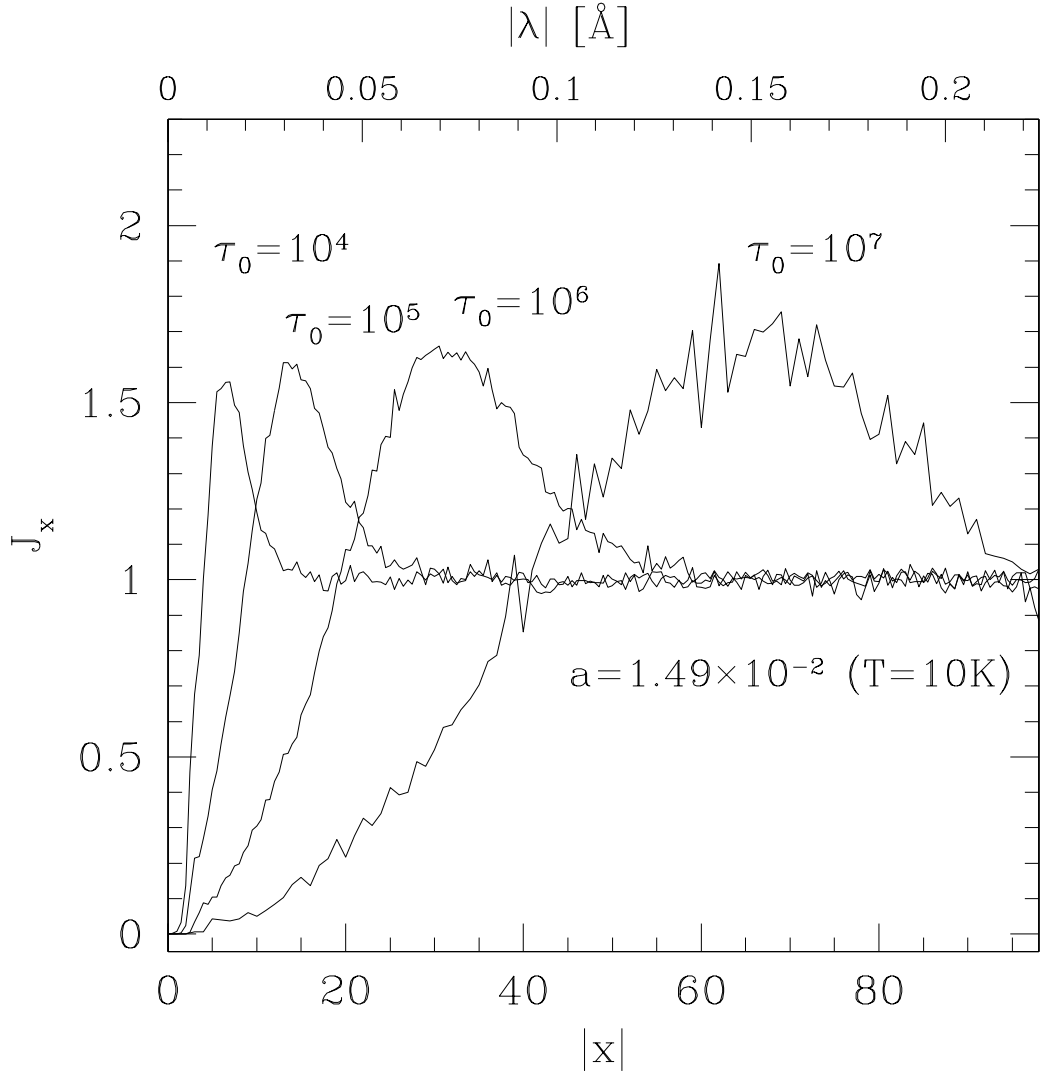}
\figcaption{
Emergent spectra near Ly$\alpha$ when the continuum
source is at the center of the scattering medium.
Here $a=1.49\times10^{-2}$ is assumed, and the spectra are normalized 
to the continuum level. We also denote the wavelength in ${\rm \AA}$ 
on the top edge of the figure.
\label{fig7}}
\end{figure}

\clearpage

\begin{figure}
\plotone{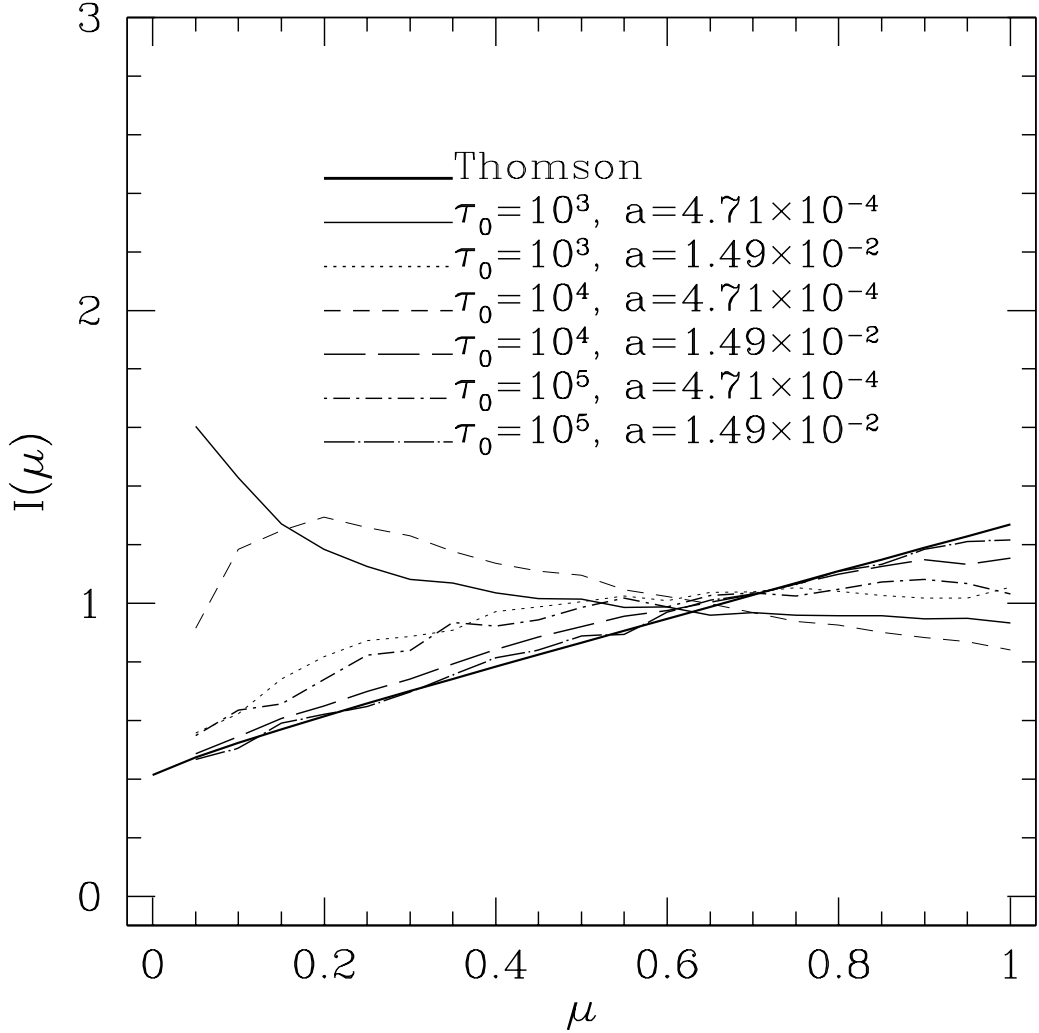}
\figcaption{
Directionality of emergent Ly$\alpha$ photons for various
optical depths $\tau_0$ and Voigt parameters $a$ of scattering media.
We define the directionality by the flux divided by $\mu$.
The solid thick line represents the limiting behavior of directionality
for the Thomson scattered radiation in an opaque electron cloud with
the radiation source deep in the cloud.
The radiation distribution is almost isotropic when the wing scattering optical
depth is small, and converges to take the form of a linear function
$I(\mu)=a+b\mu$, which is also the limit attained in the very thick Thomson
scattering medium obtained by Chandrasekhar (1960).
\label{fig8}}
\end{figure}

\clearpage

\begin{figure}
\plotone{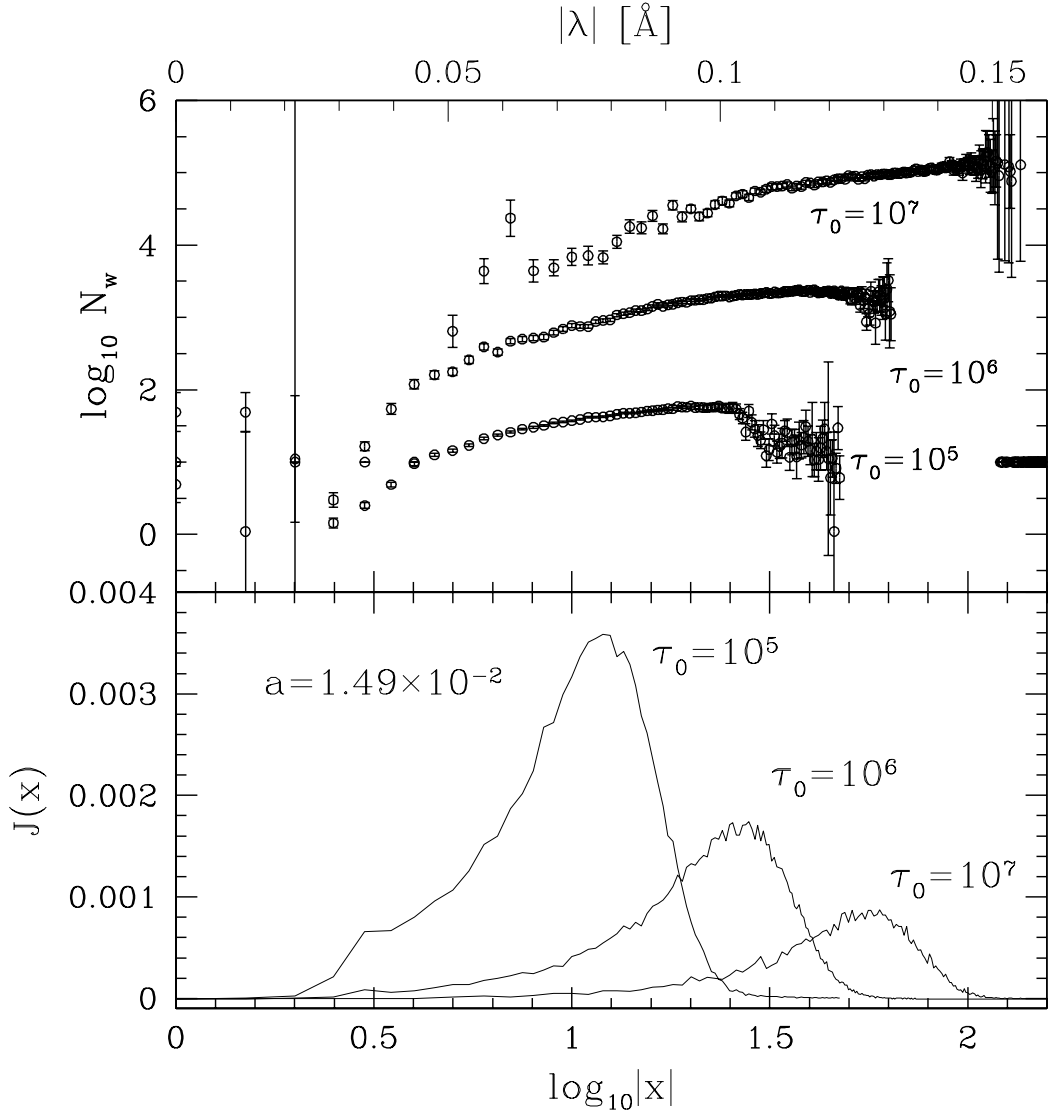}
\figcaption{Numbers of successive wing scatterings just before
escape of Ly$\alpha$ photons, $N_w$ versus the frequency, $x$.
The source is located in the midplane of the slab whose temperature
is $T=10 {\rm K}$ or $a=1.49\times10^{-2}$.
The line-center optical depths are $\tau_0=10^5, 10^6, 10^7$.
We show $N_w$ with the Poisson errors, and for clarity 
we added one and two to the curves for $\tau_0=10^6$ and $\tau_0=10^7$,
respectively.
\label{fig9}}
\end{figure}

\clearpage

\begin{figure}
\plotone{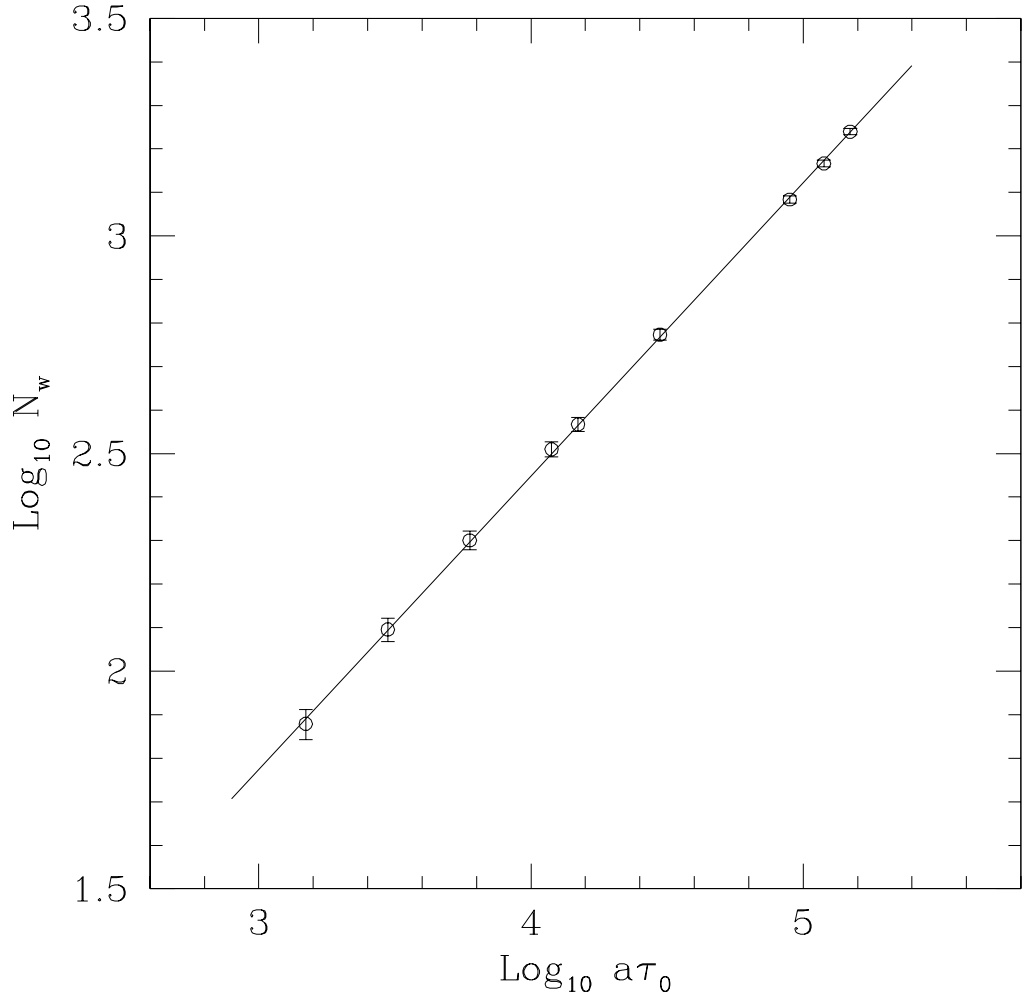}
\figcaption{
The number of a series of wing scatterings just before escape ($N_w$)
versus effective optical thickness ($a\tau_0$). Both are presented 
in logarithmic scales. The error bars are those of flux-weighted errors,
and the straight line shows the least square fit.
\label{fig10}}
\end{figure}

\end{document}